\documentclass[aps,prd,groupedaddress]{revtex4-1}

\usepackage{graphicx}
\usepackage{dcolumn}
\usepackage{bm}
\usepackage{amsmath}
 
\def\mbi#1{\mbox{\bfseries\itshape #1}}

\newcommand{\apjs}{\rm Astrophys.~J.~Supp.~}

\newcommand{\physrep}{\rm Phys.~Rep.~}

\begin{document}


\title{CMB with the background primordial magnetic field}

\author{Dai G. Yamazaki$^{1}$}
 \email{yamazaki.dai@nao.ac.jp}
\affiliation{%
$^{1}$National Astronomical Observatory of Japan, Mitaka, Tokyo 181-8588, Japan}%
\date{\today}

\begin{abstract}
We investigate the effects of the background primordial magnetic field (PMF) on the cosmic microwave background (CMB). The sound speed of the tightly coupled photon-baryon fluid is increased by the background PMF.  The increased sound speed causes the odd peaks of the CMB temperature fluctuations to be suppressed and the CMB peak positions to be shifted to a larger scale. The background PMF causes a stronger decaying potential and increases the amplitude of the CMB. These two effects of the background PMF on a smaller scale cancel out, and the overall effects of the background PMF are the suppression of the CMB around the first peak and the shifting of peaks to a large scale. We also discuss obtaining information about the PMF generation mechanisms, and we examine the nonlinear evolution of the PMF by the constraint on the maximum scale for the PMF distributions. Finally, we discuss degeneracies between the PMF parameters and the standard cosmological parameters.
\end{abstract}
\maketitle 
\section{\label{sec:introduction}Introduction}
From the discovery of magnetic fields in clusters of galaxies 
\cite{2004IJMPD..13.1549G,Wolfe:1992ab,Clarke:2000bz,Xu:2005rb} and many theoretical studies of cosmological magnetic fields\cite{Turner:1987bw,Ratra:1991bn,Bamba:2004cu, Vachaspati:1991nm,Kibble:1995aa,Ahonen:1997wh,Joyce:1997uy,Takahashi:2005nd,Hanayama:2005hd,ichiki:2006sc,Subramanian:1998fn,Mack:2001gc,Subramanian:2002nh,Lewis:2004ef,Yamazaki:2004vq,Kahniashvili:2005xe,Challinor:2005ye,Dolgov:2005ti,Gopal:2005sg,Yamazaki:2005yd,Kahniashvili:2006hy,Yamazaki:2006bq,Yamazaki:2006ah,Giovannini:2006kc,2008PhRvD..77d3005Y,Paoletti:2008ck,Yamazaki:2008bb,2008nuco.confE.239Y,Sethi:2008eq,Kojima:2008rf,2008PhRvD..78f3012K,Giovannini:2008aa,2010PhRvD..81b3008Y,2010PhRvD..81j3519Y,2010AdAst2010E..80Y,Sethi:2003vp,Sethi:2004pe,Yamazaki:2006mi,2011PhRvD..84l3006Y,2013PhRvD..88j3011Y}, it was suggested that there is a possibility of the presence of a primordial magnetic field (PMF) from the early Universe. The effect of the PMF on the early Universe and constraints on the PMF from cosmological observations are among the best-researched phenomena.

The cosmic microwave background (CMB) \cite{1965ApJ...142..419P} provides important information about the Universe. The positions of peaks and the amplitude of the CMB temperature perturbations are reflected by the sound speed of the photon-baryon fluid and the changing potential (see Ref. \cite{Dodelson:2003booka}). 
Since a magnetic field increases the sound speed of fluid and affects the evolution of density perturbations, if this amplitude as background is large enough, the magnetic field produces the critical effects on the CMB.

A power law (PL) is one of the most familiar spectra for the various physical processes including the PMF on a cosmological scale (see Refs. \cite{Grasso:2000wj,2011PhR...505....1K,2012PhR...517..141Y} and references therein).
Therefore, the effects of the PL-PMF on various physical phenomena in the Universe have been studied by many authors \cite{Grasso:2000wj,Mack:2001gc,2011PhR...505....1K,2012PhR...517..141Y,2012PhRvD..86l3006Y}.
The main parameters of the PL-PMF are the field strength on the coherent scale $\lambda$, $B_\lambda$, and the spectral index, $n_\mathrm{B}$. 
Many authors also have tried constraining these parameters from the cosmological observations \cite{2011PhR...505....1K,2012PhR...517..141Y,2012PhRvD..86l3006Y}.

Since an average of the strength of the PMF as the background is zero, while an average of the background PMF energy density is a finite value,
previous studies have constrained the background PMF energy density ($\rho_\mathrm{PMF}$) from big bang nucleosynthesis (BBN). $\rho_\mathrm{PMF}$ is proportional to the scale-invariant field strength of the PMF ($B_\mathrm{SI}$), which is as a function of $\lambda,~B_\lambda, ~n_\mathrm{B}$, and the upper and lower scales of the PMF at its generation time \cite{2012PhRvD..86l3006Y}. 
From previous studies \cite{2012PhRvD..86l3006Y}, $\rho_\mathrm{PMF}$ for larger $n_\mathrm{B}$ is comparable to the constrained PMF energy density by BBN and this influence in the CMB is not negligible. 

In this paper, we investigate the effects of the background PMF on the CMB as a function of the PL-PMF parameters: the field strength on the coherent scale, the spectral index, and the maximum scales of the PMF. We also report degeneracies of these PL-PMF parameters and the standard cosmological parameters in the PMF influences in the CMB for the first time. 

We explain how to introduce the effects of the background PMF on the CMB in Sec. II. In Sec. III, we show the equations for the numerical computation of the CMB with the PMF.
In Sec. IV, we report the PMF effects on the CMB, discuss obtaining information about the PMF generation mechanisms, and examine the nonlinear evolution of the PMF by the constraint on the maximum scale for the PMF distributions. 
We also discuss the degeneracies of the PL-PMF parameters and the standard cosmological parameters in Sec. IV.We summarize our research in Sec. V.
\section{\label{sec:models}Model}
In this section, we mention how to consider the effects of the PMF on the CMB. We modify the CAMB code \cite{camb}, which is the most familiar numerical program for computing the theoretical CMBs, taking into consideration the PMF effects.
In this paper, we use the natural units in which $\hbar = c = 1 $, where $\hbar$ is the reduced Planck constant (the Dirac constant) and $c$ is the speed of light.

From Appendix \ref{sec:app1}, the magnetic field on a scale length much bigger than $L_\mathrm{FI}$ is difficult to dissipate within the age of the Universe $t_\mathrm{age}$, and such a magnetic field is "frozen in" in the dominant fluids \cite{Dendy:1990booka}. From Eq. (\ref{eq:min}), the comoving minimum scale length of the magnetic field at last scattering is of the order of $10^{-11}$ Mpc. Thus, we can assume that the PMF is "frozen in" on the cosmological scale.
\subsection{\label{sec:modelsI}The PMF spectra}
We will introduce the power-law PMF spectra in this subsection. The detailed mathematical descriptions for them are defined in Refs.~\cite{2008PhRvD..77d3005Y,2012PhR...517..141Y}.
A power-law function is one of the most familiar spectra for distributions of the PMF \cite{Grasso:2000wj,Mack:2001gc,2011PhR...505....1K,2012PhR...517..141Y}.
A lot of authors have studied effects of PMFs with the power-law spectrum on various physical processes in the cosmology \cite{Grasso:2000wj,Mack:2001gc,2011PhR...505....1K,2012PhR...517..141Y}, 
and were challenged to constrain such PMFs.
In this paper, as in previous work, the PMF spectrum is given by the power-law.

We assume that the PMF is statistically homogeneous, isotropic, and random.
In this case, the ensemble average of the magnetic strength is zero, 
while the ensemble average of the energy density of the magnetic field, 
which is proportionate to the squared magnetic strength, has a finite value.
The PMF fluctuation spectrum can be formulated \cite{Mack:2001gc} as a power law by
$\langle B(k)B^\ast(k)\rangle \propto k^{n_\mathrm{B}}$, 
where $n_\mathrm{B}$ is the spectral index of the PMF.
We can also define a two-point correlation function of the PMF \cite{Mack:2001gc}:
\begin{eqnarray}
\left\langle B^{i}(\mbi{k}) {B^{j}}^*(\mbi{k}')\right\rangle 
	=	({(2\pi)^{n_\mathrm{B}+8}}/{2k_\lambda^{n_\mathrm{B}+3}})
		[{B^2_{\lambda}}/{\Gamma\left(\frac{n_\mathrm{B}+3}{2}\right)}]
		k^{n_\mathrm{B}}P^{ij}(k)\delta(\mbi{k}-\mbi{k}'), 
                k < k_\mathrm{max},\label{two_point1} 
\end{eqnarray}
where $P^{ij}(k)=\delta^{ij}-\frac{k{}^{i}k{}^{j}}{k{}^2}$;
$B_\lambda=|\mathbf{B}_\lambda|$ is the PMF comoving amplitude, derived by smoothing over a Gaussian sphere of radius $\lambda=1$ Mpc ($k_\lambda = 2\pi/\lambda$); and $k_\mathrm{max}$ is an upper wave number of the PMF distribution. 
Considering the nonlinear dispersion of the PMF on much smaller than cosmological scales, $k_\mathrm{max}$ is derived as a cutoff wave number $k_\mathrm{C}$ at the last scattering by Ref. \cite{Jedamzik:1996wp,Subramanian:1997gi}.

We use the PMF spectrum of the energy density $E_\mathrm{[EM:S]}(k)$, shear stress $Z_\mathrm{[EM:S]}(k)$, Lorentz forces for the scalar $\Pi_\mathrm{[EM:S]}(k)$ and vector modes $\Pi_\mathrm{[EM:V]}(k)$, and the metric source for tensor modes ($\Pi_\mathrm{[EM:T]}(k)$), that are formed by Refs.~\cite{2008PhRvD..77d3005Y,2012PhR...517..141Y}.
We also estimate them by the full numerical methods developed in our previous studies \cite{Yamazaki:2006mi,2008PhRvD..77d3005Y,2012PhR...517..141Y}.

As mentioned above, an average of the background PMF strength is zero, while an average of the background PMF energy density is a finite value.
From Ref. \cite{2012PhRvD..86l3006Y} and Appendix \ref{sec:app2}, 
the energy density and the effective amplitude of the background PMF are defined by 
\begin{eqnarray}
\rho_\mathrm{MF}
= 
\frac{1}{8\pi}
\frac{
B^2_\lambda
}
{
   \Gamma
   \left(
      \frac{n_\mathrm{B}+5}{2}
   \right)
}
\left[
(\lambda k_\mathrm{max})^{n_\mathrm{B}+3}
-
(\lambda k_\mathrm{[min]})^{n_\mathrm{B}+3}
\right],
\label{eq:BG_PL_PMF_energy_density}
\end{eqnarray}
and
\begin{eqnarray}
B_\mathrm{SI}
\equiv
B_\lambda
\sqrt{
\frac{
\left[
(\lambda k_\mathrm{[max]})^{n_\mathrm{B}+3}
-
(\lambda k_\mathrm{[min]})^{n_\mathrm{B}+3}
\right]
}
{
   \Gamma
   \left(
      \frac{n_\mathrm{B}+5}{2}
   \right)
}
}
,
\label{eq:BG_PL_PMF_amp}
\end{eqnarray}
where $\Gamma(x)$ is the gamma function, and $k_\mathrm{[min]}$ gives the minimum wave numbers and is dependent on PMF generation models. 
If the PMF is generated in the inflation epoch or produced by some vorticity anisotropies of an inflationary origin, we assume that $k_\mathrm{[min]}/k_\mathrm{max}$ is very small, and the last term of Eq. (\ref{eq:BG_PL_PMF_energy_density}) is negligibly small. The energy density and amplitude of the background PMF then reduce to 
\begin{eqnarray}
\rho_\mathrm{MF}
\sim
\frac{1}{8\pi}
\frac{
B^2_\lambda
}
{
   \Gamma
   \left(
      \frac{n_\mathrm{B}+5}{2}
   \right)
}
(\lambda k_\mathrm{max})^{n_\mathrm{B}+3}
\label{eq:BG_PL_PMF_energy_densityII}
\end{eqnarray}
and
\begin{eqnarray}
B_\mathrm{SI}
=
B_\lambda
\sqrt{
\frac{
(\lambda k_\mathrm{max})^{n_\mathrm{B}+3}
}
{
   \Gamma
   \left(
      \frac{n_\mathrm{B}+5}{2}
   \right)
}
}
.
\label{eq:BG_PL_PMF_ampII}
\end{eqnarray}
\subsection{\label{sec:modelsII}Cosmic expansion rate with background PMF energy density}
In this subsection, we introduce the effect of the background PMF on the cosmic expansion. 
In a homogeneous and isotropic flat universe, we can find the Hubble parameter $H$ from 
the Friedmann equation and the conservation of energy momentum-tensor as follows;
\begin{eqnarray}
\left(
\frac{\dot{a}}{a}
\right)^2
 \equiv H^2 = \frac{8\pi G}{3}\rho \label{eq_friedmann},\\
\dot{\rho} = -3H
\left(
	\rho + p
\right)
\label{eq_conservation},
\end{eqnarray}
where 
overdots represent derivatives with respect to time, 
$G$ is Newton's constant, $\rho$ is the total energy density, and $p$ is the total pressure.
We assume that the main fluid components in the Universe are the photon, 
the neutrino, cold dark matter (CDM), and the baryon.
In this paper, variables with the subscripts ``$\gamma$", ``$\nu$", ``CDM" and ``b" indicate the photon, neutrino, CDM, and baryon, respectively. 
Therefore, the total energy density and pressure with the background PMF are 
$\rho = \rho_\gamma + \rho_\nu + \rho_\mathrm{CDM}+ \rho_\mathrm{b} + \rho_\mathrm{MF}$
and
$p = p_\gamma + p_\nu + p_\mathrm{MF}$,
where $\rho_\mathrm{MF}$ and $p_\mathrm{MF}$ are the energy density and pressure of the background PMF, and  we consider the matter pressures to be much smaller than the radiation ones.

If $\rho_\mathrm{MF}$ is large enough in the radiation-dominated era, the effect of the energy density of the background PMF on the cosmic expansion is not small, and the matter-radiation equality time $t_\mathrm{eq}$becomes later.
In this case, the decaying potential by the radiation becomes relatively strong on larger scales at the recombination, and it makes the forced oscillation of the photon-baryon fluid and the early integrated Sachs-Wolfe (ISW) effect stronger.
Therefore, a sufficiently large energy density of the PMF in the radiation-dominated era has no small effect on the primary temperature fluctuations of the CMB for lower $\ell$, and we can expect that the PMF energy density can be constrained by the CMB temperature fluctuations for lower $\ell$.
\subsection{\label{sec:app3} A magnetosonic wave}
In this subsection, we will introduce the magnetohydrodynamics (MHD) equations and derive a magnetic sonic speed considering the background PMF.

The MHD equations\cite{1992pavi.book.....S} are
\begin{eqnarray}
\rho
\left(
     \frac{\partial}{\partial t}
   + \mbi{u}\cdot \nabla
\right)\mbi{u}
&=&
-\nabla p
+\frac{1}{4\pi}
\left(
   \nabla \times \mbi{B}
\right)\times \mbi{B}
\label{eq_motion_mhd}\\
\frac{\partial}{\partial t}\rho
&=&
-\nabla\cdot(\rho\mbi{u})
\label{eq_mx1_mhd}\\
\frac{\partial}{\partial t}\mbi{B}
&=&
\nabla \times (\mbi{u}\times\mbi{B})
\label{eq_mx2_mhd}
\end{eqnarray}
We assume $\rho=\rho_0$, $p=p_0$, $\mbi{u}=0$, and $\mbi{B}=\mbi{B}_0$ as the steady state and $\delta\rho$, $\delta\mbi{u}$, $\delta p$, $\delta \mbi{B}$ as the small perturbations.
The first-order perturbations of MHD equations then are
\begin{eqnarray}
\rho_0
     \frac{\partial}{\partial t}
     \delta \mbi{u}
&=&
-\nabla \delta p
+\frac{1}{4\pi}
   \nabla \times \
(\delta\mbi{B} \times \mbi{B})
\label{eq_motion_mhdP}\\
\frac{\partial}{\partial t}\delta\rho
&=&
-\nabla\cdot(\rho_0\delta\mbi{u})
\label{eq_mx1_mhdP}\\
\frac{\partial}{\partial t}\delta\mbi{B}
&=&
\nabla \times (\delta\mbi{u}\times\mbi{B})
\label{eq_mx2_mhdP}\\
\delta p
&=&
c^2_s \delta \rho,
\label{eq_sound_speed}
\end{eqnarray}
where $c_s$ is a sound speed without a magnetic field.
We assume the oscillation of the first-order perturbation to be $\exp{[i(\mbi{k}\cdot\mbi{x}-\omega t)]}$; we can use the following relations;
$\partial / \partial t \rightarrow -i\omega$, $\nabla \rightarrow i\mbi{k}$, and $\nabla \times \rightarrow i\mbi{k}\times$.
Therefore, using Eq. (\ref{eq_sound_speed}), Eqs. (\ref{eq_motion_mhdP})-(\ref{eq_mx2_mhdP}) become
\begin{eqnarray}
\omega\rho_0\delta \mbi{u}
&=&
c^2_s
\delta \rho
\mbi{k}
-\frac{1}{4\pi}
   (\mbi{k} \times \delta\mbi{B}) \times \mbi{B})
\label{eq_motion_mhdPf}\\
\omega\delta\rho
&=&
\rho_0(\mbi{k}\cdot\delta\mbi{u})
\label{eq_mx1_mhdPf}\\
\omega\delta\mbi{B}
&=&
-\mbi{k} \times (\delta\mbi{u}\times\mbi{B})
\label{eq_mx2_mhdPf}
\end{eqnarray}
Using Eqs. (\ref{eq_mx1_mhdPf}) and (\ref{eq_mx2_mhdPf}), 
Eq. (\ref{eq_motion_mhdPf}) then becomes
\begin{eqnarray}
\omega^2\delta \mbi{u}
&=&
c^2_s
(\mbi{k}\cdot\delta \mbi{u})
\mbi{k}
+\frac{1}{4\pi\rho_0}
   \{\mbi{k} \times
      [\mbi{k} \times
         (\delta \mbi{u} \times \mbi{B}) 
       ]
    \}
\times \mbi{B}
\label{eq_motion_mhdPfII}
\end{eqnarray}
Using 
$\mbi{A}\times(\mbi{B}\times\mbi{C}) = 
(\mbi{A}\cdot\mbi{C})\mbi{B}
-
(\mbi{A}\cdot\mbi{B})\mbi{C}
$
and
$(\mbi{A}\times\mbi{B})\times\mbi{C} = 
(\mbi{A}\cdot\mbi{C})\mbi{B}
-
(\mbi{B}\cdot\mbi{C})\mbi{A}$
, the last term of Eq. (\ref{eq_motion_mhdPfII}) is
\begin{eqnarray}
\frac{1}{4\pi\rho_0}
   \{\mbi{k} \times
      [\mbi{k} \times
         (\delta \mbi{u} \times \mbi{B}) 
       ]
    \}
\times \mbi{B}
=
\nonumber \\
\frac{1}{4\pi\rho_0}
\left[
     (\mbi{k}\cdot\mbi{B})^2 \delta \mbi{u}
   - (\mbi{k}\cdot\mbi{B})(\delta \mbi{u}\cdot\mbi{B})\mbi{k}
   - (\mbi{k}\cdot\delta \mbi{u})(\mbi{k}\cdot\mbi{B})\mbi{B}
   + (\mbi{k}\cdot\delta \mbi{u})B^2\mbi{k}
\right]
\label{last_term_eq_motion_mhdPfII}
\end{eqnarray}
We assume that the direction of the magnetic field is fixed as previous work and the texts \cite{1992pavi.book.....S,Adams:1996cq} indicate so far.
In statistical cosmological study, we are interested not in local magnetic effects but in global ones, which are average universe-wide.
We also assume that the background PMF is stochastic isotropic and homogenous.
In this case, the ensemble average of the magnetic strength is zero, 
while the ensemble average of the energy density of the magnetic field has a finite value.
Therefore, the relation between the mean square of the background PMF amplitude $\langle B_\mathrm{bc}^2 \rangle  = B^2_\mathrm{SI} \propto \rho_\mathrm{MF}$ and each spatial component $\langle B_x^2 \rangle $, $\langle B_y^2 \rangle $ and $\langle B_z^2 \rangle $ is $\langle B_x^2 \rangle = $ $\langle B_y^2 \rangle =$ $\langle B_z^2 \rangle =$ $\frac{1}{3}B^2_\mathrm{SI}$.
From these interpretations, in Eq. (\ref{last_term_eq_motion_mhdPfII}), 
the average of $\mbi{k}\cdot\mbi{B} = kB\cos \theta $ per $\theta$ becomes zero, and 
the average of $(\mbi{k}\cdot\mbi{B})^2 = k^2B^2(\cos \theta)^2 $ per $\theta$ becomes 
$k^2 B^2_\mathrm{SI}/6$.
Then, we obtain
\begin{eqnarray}
\omega^2\delta \mbi{u}
&=&
c^2_s
(\mbi{k}\cdot\delta \mbi{u})
\mbi{k}
+
\frac{B^2_\mathrm{SI}}{12\pi\rho_0}
\left[
     \frac{1}{2}k^2  \delta \mbi{u}
   + (\mbi{k}\cdot\delta \mbi{u})\mbi{k}
\right].
\label{eq_motion_mhdPfIII}
\end{eqnarray}
From the inner product of Eq. (\ref{eq_motion_mhdPfIII}) and $\mbi{k}$, 
the effective sound speed with the background PMF is derived by
\begin{eqnarray}
c^2_\mathrm{sA}
=
c^2_\mathrm{s}
+
\frac{1}{2}
c^2_\mathrm{A},
\label{result_c_sA}
\end{eqnarray}
where $c^2_\mathrm{A}$ is the Alfven speed from the background PMF, defined by $ c^2_\mathrm{A} = \frac{\langle B_\mathrm{bc}^2 \rangle}{4\pi\rho_0}$.
\section{Equations with the primordial magnetic field}
In this section, we will introduce the essential evolution equations with the PMF 
for each mode.
For details, see Refs.\cite{2008PhRvD..77d3005Y,Shaw:2009nf}.
In this paper, we choose the conformal Newtonian gauge, that is defined by Refs. \cite{Ma:1995ey,Hu:1997hp,Shaw:2009nf}. 
After this section, we also use the conformal time, which is defined by $\int^t_0 dt'/a(t')$, instead of physical time, $t$.
In the linear approximation, we should consider the zero-order factor, e.g. radiation, matter, and background PMF energy densities, for solving all equations as mentioned in SubSec. \ref{sec:modelsII}, while solutions of perturbations can be divided into those with and without PMF as the first-order perturbation by Green's function method as, 
\begin{eqnarray}
f(k)= f_\mathrm{[FL]}(k)+f_\mathrm{[PMF]}(k).
\label{correlation_f}
\end{eqnarray}
In this paper, we assume that there is no correlation between the PMF and the primary perturbations. In this case, we do not have to consider the correlation term between the PMF and primary in Eq. (\ref{correlation_f}). 
Since the equations for $f_\mathrm{[FL]}(k)$ are equal to the equations for $f_\mathrm{[PMF]}(k)$ which are removed the PMF terms as the first order perturbation source except the background PMF, we do not write down the equations for $f_\mathrm{[FL]}(k)$ in this paper.
From 
Refs.\cite{Padmanabhan:1993booka,
Ma:1995ey,
Hu:1997hp,
Hu:1997mn,
Dodelson:2003booka,
Giovannini:2006gz,
2008PhRvD..77d3005Y,
Shaw:2009nf}, 
the evolution equations of the scalar mode with the PMF are
\begin{eqnarray}
k^2\phi + 3H(\dot{\phi}+H\psi) &=& 4\pi G{a^2}
\left\{
	E_\mathrm{[EM:S]}(\mbi{k},\tau)-\delta\rho
\right\} \label{eq:phi}\\
k^2(\phi-\psi) &=& 
-12\pi G{a^2}
\left\{
	Z_\mathrm{[EM:S]}(\mbi{k},\tau)
	-(\rho_\nu+P_\nu)\sigma_\nu
	-(\rho_\gamma+P_\gamma)\sigma_\gamma
\right\}
 \label{eq:phi_psi}\\
\dot{\delta}^\mathrm{(S)}_\mathrm{CDM}
 	&=&
 		-v^\mathrm{(S)}_\mathrm{CDM}+3\dot{\phi}~,\label{eq:CDM_rho}\\
\dot{v}^\mathrm{(S)}_\mathrm{CDM}
 	&=&
        -\frac{\dot{a}}{a}v^\mathrm{(S)}_\mathrm{CDM}+k^2\psi~,\label{eq:CDM_v}\\
\dot{\delta}^\mathrm{(S)}_{\gamma}
 	&=&
 		-\frac{4}{3}v^\mathrm{(S)}_{\gamma}
 		+4\dot{\phi}~,\label{eq:photon_rho}\\
\dot{\delta}^\mathrm{(S)}_{\nu}
 	&=&
 		-\frac{4}{3}v^\mathrm{(S)}_{\nu}
 		+4\dot{\phi}~,\label{eq:nu_rho}\\
\dot{v}^\mathrm{(S)}_{\gamma}
	&=&
		k^2\left(\frac{1}{4}\delta^\mathrm{(S)}_{\gamma}-\sigma_{\gamma}\right)
		+an_e\sigma_T(v^\mathrm{(S)}_\mathrm{b}-v^\mathrm{(S)}_{\gamma})~+k^2\psi,
		\label{eq:photon_v} \\
\dot{v}^\mathrm{(S)}_{\nu}
	&=&
		k^2\left(\frac{1}{4}\delta^\mathrm{(S)}_{\nu}-\sigma_{\nu}\right)+k^2\psi,\label{eq:nu_v} \\
\dot{\delta}^\mathrm{(S)}_\mathrm{b}
	&=&
		-v^\mathrm{(S)}_\mathrm{b}+3\dot{\phi}
 			\label{eq:baryon_rho}  \\
\dot{v}^\mathrm{(S)}_\mathrm{b}
	&=&
			-\frac{\dot{a}}{a}v^\mathrm{(S)}_\mathrm{b}
	+c^2_\mathrm{bA}k^2\delta^\mathrm{(S)}_\mathrm{b}
                        +\frac{1}{R} an_e\sigma_T(v^\mathrm{(S)}_{\gamma}-v^\mathrm{(S)}_\mathrm{b})+k^2\psi
\nonumber\\
&&			
			+
			k^2\frac{\Pi_{\mathrm{[EM:S]}}(\mbi{k},\tau)}{\rho_b},
			\label{eq:baryon_v}
\end{eqnarray} 
where 
$\psi$ is the perturbation of gravitational potential in the Newtonian limit;
$\phi$ is the perturbation of the spatial curvature;
$R$ is $(3/4)(\rho_b/\rho_\gamma)$;
$\sigma_{\gamma}$ and $\sigma_{\nu}$ are the shear stresses of the photons and the neutrino, respectively; 
$v^\mathrm {(S)}_X$ and $\delta^\mathrm {(S)}_X$ are the velocity and density perturbations for each component $X$;
$n_e$ is the free electron number density,;
and
$\sigma_T$ is the Thomson scattering cross section.
Here $c^2_\mathrm{bA}=c^2_\mathrm{b} + c^2_\mathrm{A}/2$ is the magnetosonic speed in baryon fluid [see Eq. (\ref{result_c_sA})], where $ c^2_\mathrm{b}$ is the sound speed of the baryon fluid without the background PMF  and is defined by Ref. \cite{Ma:1995ey}.

The evolution equations of the vector mode with the PMF are
\begin{eqnarray}
k
\left(
	\dot{V}+2\frac{\dot{a}}{a}V
\right)
=
- 8\pi a^2 G
\left[
2 \Pi_\mathrm{[EM:V]} (\mathbf{k},\tau)
+ p_\gamma\pi_\gamma+p_\nu\pi_\nu
\right]
\label{eq:vectorpoten}
\\
\dot{v}_{\nu}^\mathrm{(V)}-\dot{V}
	= -k\left(\frac{\sqrt{3}}{5}\Theta^\mathrm{(V)}_{\nu 2}\right),
\label{eq:vneutrino1}\\
 \dot{v}_{\gamma}^\mathrm{(V)}
-\dot{V}
+\dot{\tau}_c
    (
	v^\mathrm{(V)}_{\gamma}-v^\mathrm{(V)}_{b}
	)
=-k\left(
			\frac{\sqrt{3}}{5}\Theta^\mathrm{(V)}_{\gamma 2}
   \right),
\label{eq:vphoton1}\\
 \dot{v}_{b}^\mathrm{(V)}
-\dot{V}
+\frac{\dot{a}}{a}(v_{b}^\mathrm{(V)}-V)
-\frac{1}{R}\dot{\tau}_c(v^\mathrm{(V)}_{\gamma}-v^\mathrm{(V)}_{b})\nonumber \\
	=
		k\frac{\Pi_\mathrm{[EM:V]} (\mathbf{k},\tau)}{\rho_b},
\label{eq:vbaryon1}
\end{eqnarray}
where $V(\tau, \mathbf{k})$ is the vector potential;
$p_X$, $\pi_X $, $v^{(V)}_X$ are the pressure, the anisotropic stress, and velocity for each component $X$; 
$\Theta^{(V)}_{\nu 2}$ and $\Theta^{(V)}_{\gamma 2}$
 are quadrupole moments of the neutrino and photon angular distributions, respectively \cite{Hu:1997hp,Hu:1997mn}.

The evolution equations of the tensor mode with the PMF are
\begin{eqnarray}
\ddot{\mathcal{H}}
+2\frac{\dot{a}}{a}\dot{\mathcal{H}}
+k^2\mathcal{H}=
8\pi G a^2 
	\left(
		\Pi_\mathrm{[EM:T]}+\frac{8}{5}\Theta_2^{(T)}
	\right), 
\end{eqnarray}
where $\mathcal{H}$ is the tensor potential and
$\Theta_2^{(T)}$ is the quadrupole moment of the photon angular distribution \cite{Hu:1997hp,Hu:1997mn}.

We shall explain the effects of the vector and tensor modes with the background PMF on the CMB. 
The vector and tensor modes do not have terms which are directly dependent on the oscillatory propagations, as the second term on the left side of Eq. (\ref{eq:baryon_v}). These modes also do not have terms which are directly dependent on energy-density perturbations.
Since it is difficult for the radiation-like energy densities including the PMF to contribute to the expansion rate of the Universe around the epoch of the recombination, the effect of the PMF energy density on $H$ is subdominant. Thus, the background PMF effects on the vector and tensor mode are relatively small.

Since the sonic speed of the baryon fluid with the background PMF is not an effective factor for the mentioned equations at the subhorizon and superhorizon, for deriving initial conditions, we do not have to consider $c_\mathrm{bA}$, and it is only necessary to consider the background PMF for 
Eqs. (\ref{eq_friedmann}) and (\ref{eq_conservation}).
Therefore, the values of the initial conditions are dependent on the total energy density and Eqs. (\ref{eq_friedmann}) and (\ref{eq_conservation}), which are affected by the background PMF, and 
we just have to change $\rho$ and $p$ without the background PMF to their values with the background PMF.
Finally, we do not have to change the expression of the initial conditions in Refs. \cite{2008PhRvD..77d3005Y,Shaw:2009nf}.
\section{\label{sec:results}Results and Discussions}
As mentioned in Sec. \ref{sec:models}, since there is no density perturbation term in the equations of the vector and tensor modes, we have no need to consider the sonic speed term for these modes as the second term on the left side of Eq.(\ref{eq:baryon_v}).
Therefore, at first, to understand the pure background PMF effect on the CMB from the scalar mode, we derive the analytical solution from the adiabatic initial condition for the acoustic oscillation of the photon-baryon fluid without the PMF term as the first-order perturbation, which is the last term of Eq. (\ref{eq:baryon_v}), in the scalar mode.

From Eqs. (\ref{eq:phi}), (\ref{eq:phi_psi}), (\ref{eq:photon_rho}), (\ref{eq:photon_v}), (\ref{eq:baryon_rho}), and (\ref{eq:baryon_v}) without the last term, we obtain as
\begin{eqnarray}
\ddot{\delta}^\mathrm{(S)}_\gamma+
   H\frac{R}{1+R}\dot{\delta}^\mathrm{(S)}_\gamma + 
k^2c_\mathrm{S}^2\delta^\mathrm{(S)}_\gamma
=
   \ddot{\phi} - \frac{R}{1+R}\dot{a}\dot{\phi} - \frac{k^2}{3}\psi.
\label{eq1}
\end{eqnarray}
Here $c^2_\mathrm{S} = c^2_\mathrm{pb} + c^2_\mathrm{A}/2$, where $c^2_\mathrm{pb} = \frac{1}{3(1+R)}$ is the sound speed of the photon-baryon fluid without the background PMF, and $c^2_\mathrm{A}$ is derived using the cutoff scale derived by the nonlinear dispersion model \cite{Jedamzik:1996wp,Subramanian:1997gi}.
On the right side of Eq. (\ref{eq1}), the first term is a time delay from the Universe expansion, the second term is an effect of the Universe expansion, and the third term is a blue (or red) shift from the gravity potential.
Since we are interested in phenomena of the photon-baryon fluid around the epoch of the recombination, we consider the matter-dominant era.
In the matter-dominant era, 
the potential terms are not dependent on time before the cosmological constant dominates the Universe. Therefore, we can neglect the first and second terms of the right side of Eq. (\ref{eq1}), and the third term also is not dependent on time.
Finally, we obtain
\begin{eqnarray}
\ddot{\delta}^\mathrm{(S)}_\gamma+
   H\frac{R}{1+R}\dot{\delta}^\mathrm{(S)}_\gamma + 
k^2c_\mathrm{S}^2\delta^\mathrm{(S)}_\gamma
\sim
- \frac{k^2}{3}\psi.
\label{eq2}
\end{eqnarray}
The special solution of Eq. (\ref{eq2}) is 
\begin{eqnarray}
\delta^\mathrm{(S)}_\gamma
=
- \frac{1}{3 c_\mathrm{S}^2}\psi.
\label{ans1}
\end{eqnarray}
In case of the adiabatic condition, the homogeneous solution of Eq. (\ref{eq2}) is
\begin{eqnarray}
\delta^\mathrm{(S)}_\gamma
=
   A\cos{(kd_\mathrm{S})},
\label{ans2}
\end{eqnarray}
where $ d_\mathrm{S}$ is the sound horizon as $\int^\eta_0 c_\mathrm{S}(\eta') d \eta' $.
From the large-scale limit of the Boltzmann equation, the initial condition of $\delta^\mathrm{(S)}_\gamma$ in the matter-dominant epoch is $-2\psi(0)/3$.
Therefore, from Eqs. (\ref{ans1}) and (\ref{ans2}), 
the general solution of Eq. (\ref{eq2}) is 
\begin{eqnarray}
\delta^\mathrm{(S)}_\gamma (\eta)
=
     \left[\frac{1}{3c_\mathrm{S}^2} - \frac{2}{3}\right]
     \psi
     \cos{(kd_\mathrm{S})}
   - \frac{1}{3c_\mathrm{S}^2}\psi.
\label{ansF}
\end{eqnarray}
An observable which we can obtain from the CMB is $\delta^\mathrm{(S)}_\gamma (\eta) + \psi$, because the temperature fluctuations of the CMB are affected by the gravitational redshift from the gravitational potential. We also assume the gravitational potential to be an external field, and a gravitational potential has a negative value if a density fluctuation has a positive value. 
Taking into account these considerations, and from Eq. (\ref{ansF}), we therefore show that the observable $\delta^\mathrm{(S)}_\gamma (\eta) + \psi$ is
\begin{eqnarray}
|\delta^\mathrm{(S)}_\gamma (\eta) + \psi|
=
\left[
     \left(
\frac{2}{3} - \frac{1}{3c_\mathrm{S}^2} 
\right)
\cos{(kd_\mathrm{S})}
   \left(\frac{1}{3c_\mathrm{S}^2} -1 \right)
\right] 
|\psi|.
\label{eq_AO}
\end{eqnarray}
From this equation, the amplitudes of odd peaks decrease when the sound speed increases, while the amplitudes of even peaks are not affected by the increased sound speed. Furthermore, the wavelength of Eq. (\ref{eq_AO}) increases when the sound speed increases, and the peak positions are shifted to a larger scale.

\begin{figure}
\includegraphics[width=1.0\textwidth]{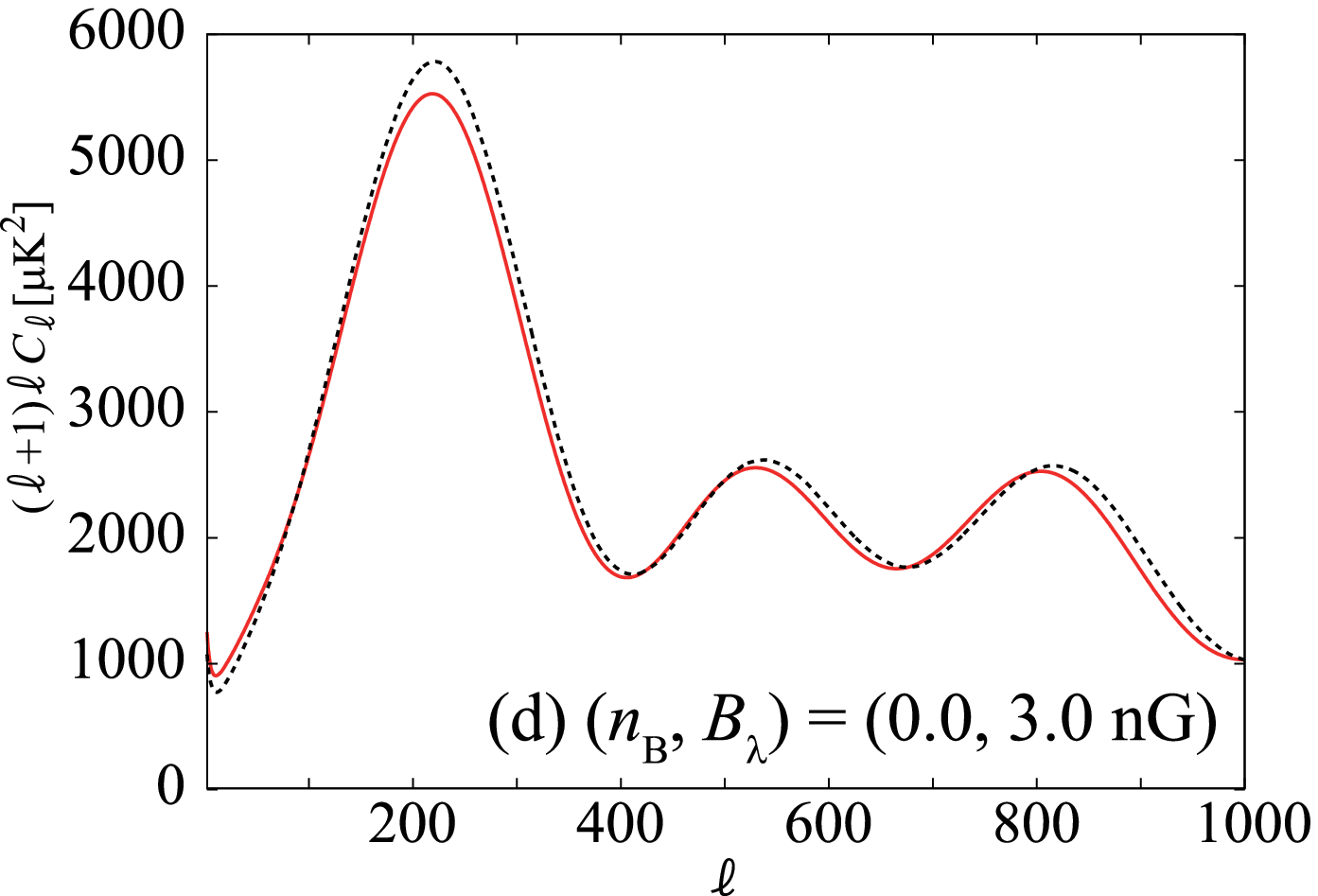}
\caption{\label{fig2}
The effects of the background magnetic field on the CMB.
The dotted curve is the theoretical result from the WMAP nine-year best-fit parameter in $\Lambda$CDM and the tensor mode\cite{WMAP_9yr_Arxiv} \footnote{The main papers of the Planck project are under review, so we refrain from using their results}.
These standard cosmological parameters are $(\Omega_b, ~\Omega_\mathrm{CDM},~n_s,~ 10^9\Delta^2_R, ~H_0, ~\tau, r) = (0.0442, ~0.210,~0.992,~2.26,~72.6,~0.091,~0.38)$ , 
where 
$\Omega_b h^2$ is the baryon density,
$\Omega_\mathrm{CDM} h^2$ is the CDM density,
$n_s$ is the scalar spectral index,
$10^9\Delta^2_R$ is the amplitude of the initial fluctuation, 
$H_0$ is the Hubble parameter,
$\tau$ is the optical depth, and
$r$ is the tensor-to-scalar ratio.
The bold curve is the theoretical result with the background PMF effects of $(0.0,~3~\mathrm{nG})$ (without the first-order perturbation of the PMF). In this case, $\rho_\mathrm{MF}/\rho_\gamma = 0.0161$.
} 
\end{figure}

If we consider the energy density of the background PMF, the total radiation-like energy density $\rho_R$ increase and the epoch of equality occurr closest to the recombination, so that $\rho_R$ has to be accounted for in estimating the temperature fluctuations of the CMB at the recombination. In this case, the decaying potential by the radiation becomes relatively strong on larger scales at the recombination, and it provides a stronger driving force for the oscillations and the stronger early ISW effect. Thus, the amplitude of the CMB is larger than in a universe without the background PMF. Since the decaying potentials occur in the horizon, and the smaller scales enter the horizon earlier, the potentials of larger scales decay more weakly. Therefore, the increase in the amplitude from these effects is smaller around the first peak, and this effect also cancels out the effects of $c_\mathrm{S}$ on the odd peaks less than the third peak of the CMB.
Finally, the total changing amplitudes of the CMB from the pure effects of the background PMF around the first peak are stronger than in smaller scales.
Actually, these features are illustrated by theoretical computed results of CMB temperature fluctuations with the background PMF in Fig. \ref{fig2} \footnote{Note: in this figure, we do not consider the last term of Eq. (\ref{eq:baryon_v})}.

\begin{figure}
\includegraphics[width=1.0\textwidth]{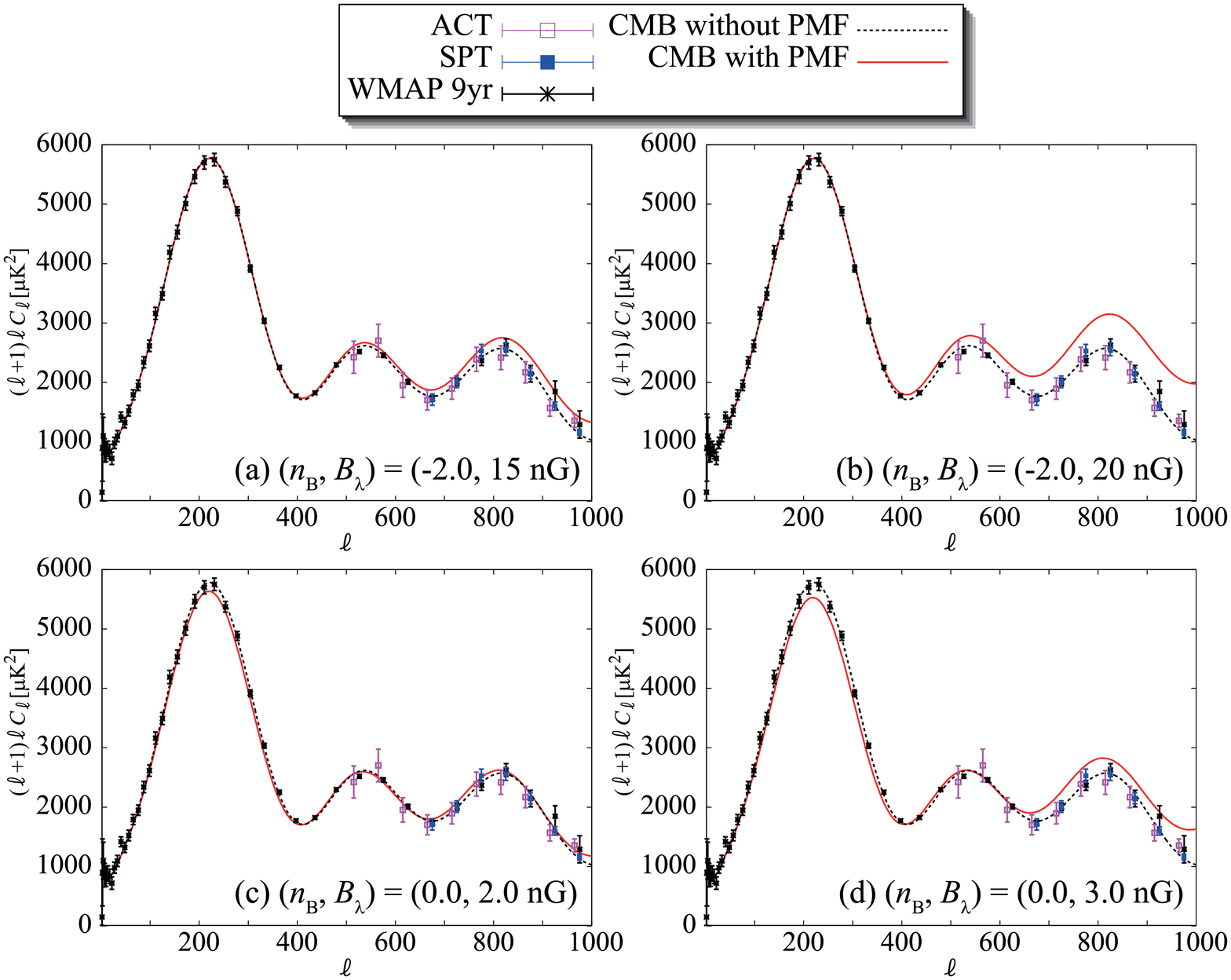}
\caption{\label{fig3}
The effects of the PMF on the CMB.
The dotted curve is the theoretical result from the WMAP nine-year best-fit parameter in $\Lambda$CDM and the tensor mode\cite{WMAP_9yr_Arxiv}\footnote{The main papers of the Planck project are under review, so we refrain from using their results}.
The bold curves in panels (a), (b), (c) and (d) are the theoretical results with the PMF effects of $(n_\mathrm{B},~B_\lambda,~\rho_\mathrm{MF}/\rho_\gamma)~=~(-2.0,~15~\mathrm{nG},~0.00116),~(-2.0,~20~\mathrm{nG},~0.00185),~(0.0,~2~\mathrm{nG}, ~0.00910)$, and $(0.0,~3~\mathrm{nG},~0.0161)$, respectively.
The dots with the error bars are the results of the CMB observations\cite{WMAP_9yr_Arxiv,2011ApJ...739...52D,2011ApJ...743...28K}, as shown by the legend in this figure.
} 
\end{figure}

\begin{figure}
\includegraphics[width=1.0\textwidth]{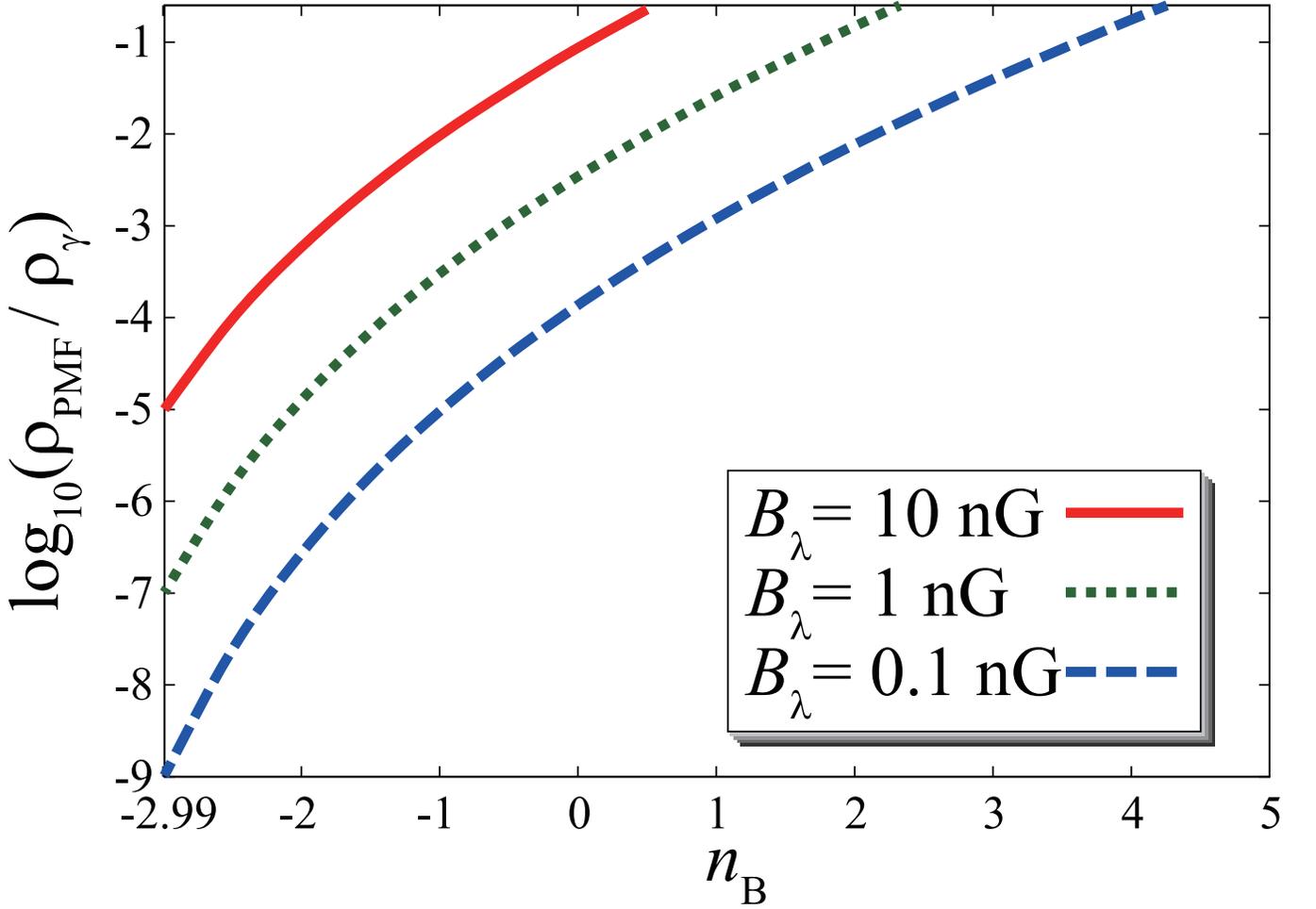}
\caption{\label{fig1}
The ratio of the background magnetic field to photon density.
The bold, dotted, and dashed curves are $B_\lambda = 10, 1,$ and $0.1$ nG, respectively. 
} 
\end{figure}

Considering these effects of the background PMF on the CMB, we can effectively constrain them by the observation results on lower $\ell$. 
Therefore, next, we shall focus on the effect of the background PMF on the CMB for $\ell < 1000$.
Figure. \ref{fig3} shows the temperature fluctuations of the CMB with total PMF effects (scalar + vector + tensor modes and the background PMF).
From Eq. (\ref{eq:BG_PL_PMF_energy_densityII}) and Fig. \ref{fig1}, the energy density of the background PMF is dependent on the power-law index $n_\mathrm{B}$ and becomes much smaller with lower $n_\mathrm{B}$. 
Therefore, the effects of the PMF at lower $n_\mathrm{B}$ are dominated by the perturbation-like PMF as in previous studies. 
Actually, in Figs. \ref{fig3}(a) and \ref{fig3}(b), the effects of the background PMF on the CMB around the first peak are very small, even if the $B_\lambda$'s are larger than the previous constrained  values. 
On the other hand, the effects of the PMF of bigger $n_\mathrm{B}$ on the CMB are not negligible around the first peak (Figs. \ref{fig3}(c) and \ref{fig3}(d)).
Since, the observational result of the CMB around the first peak is much better than at higher peaks, we expect that the PMF of the nonlinear cutoff model on bigger $n_\mathrm{B}$ can be constrained more strongly.

\begin{figure}
\includegraphics[width=1.0\textwidth]{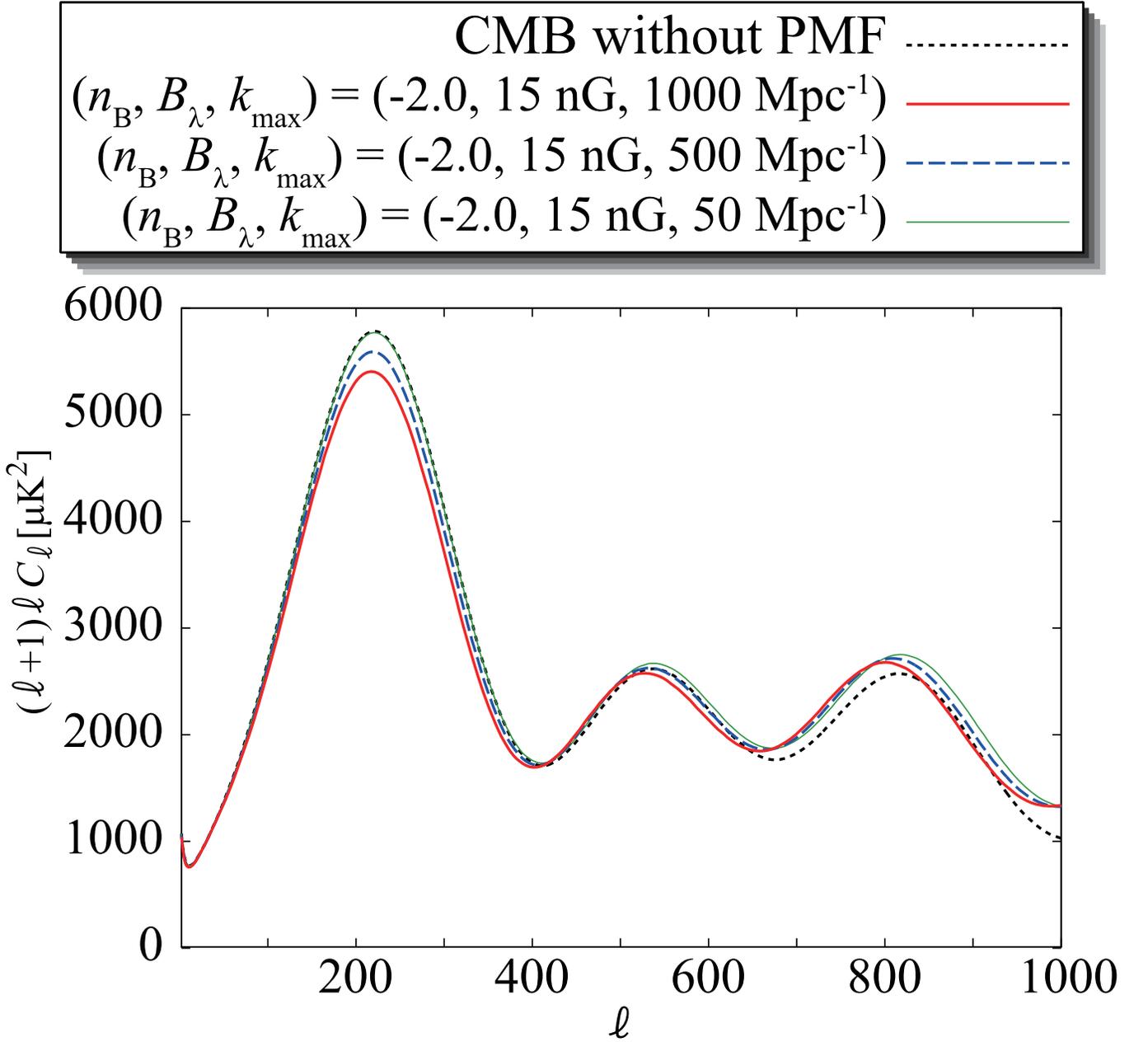}
\caption{\label{fig4}
The contribution of the upper-limit wave number of the PMF to the CMB. The dotted curve is the theoretical result from the WMAP nine year best-fit parameter in $\Lambda$CDM and the tensor mode\cite{WMAP_9yr_Arxiv}.
The bold, dashed, and thin curves are the theoretical results with the PMF effects of $(n_\mathrm{B},~B_\lambda,~k_\mathrm{max},~\rho_\mathrm{MF}/\rho_\gamma)~=~(-2.0,~15~\mathrm{nG}, ~1000~\mathrm{Mpc}^{-1},~0.0240),~(-2.0,~15~\mathrm{nG}, ~500~\mathrm{Mpc}^{-1},~0.0120)$,and $(-2.0,~15~\mathrm{nG}, ~50~\mathrm{Mpc}^{-1},~0.00120)$,  respectively. 
} 
\end{figure}

We shall discuss $k_\mathrm{max}$ being assumed as a free parameter as an academic interest.
Mathematically, in this case, the energy density of the PMF is dependent on the wave-number upper limit $k_\mathrm{max}$ [Eq. (\ref{eq:BG_PL_PMF_energy_density})], and $k_\mathrm{max}$ is dependent on a generation mechanism of the PMF. Using this property, we can obtain the prior limits of $k_\mathrm{max}$ and constrain PMF generation models indirectly from the CMB.
Figure \ref{fig4} shows the contribution of $k_\mathrm{max}$ to the CMB. A larger $k_\mathrm{max}$ induces larger $\rho_\mathrm{MF}$ and $c_\mathrm{S}$. Thus, the locations of the peaks and troughs of the CMB are shifted to smaller $\ell$, and the amplitude of the CMB around the first peak is suppressed. This change is very unique, and we expect that $k_\mathrm{max}$ can be constrained by the CMB of lower $\ell$. The constraint on $k_\mathrm{max}$ also helps to determine whether or not the diffusion model of the PMF on the nonlinear region is plausible.
If the large-$k_\mathrm{max}$ model has a better likelihood, we should construct a new physical model for the PMF time evolution in the nonlinear region; on the other hand, we will be able to confirm that the previous model that derives $k_C$ is suitable.

\begin{figure}
\includegraphics[width=1.0\textwidth]{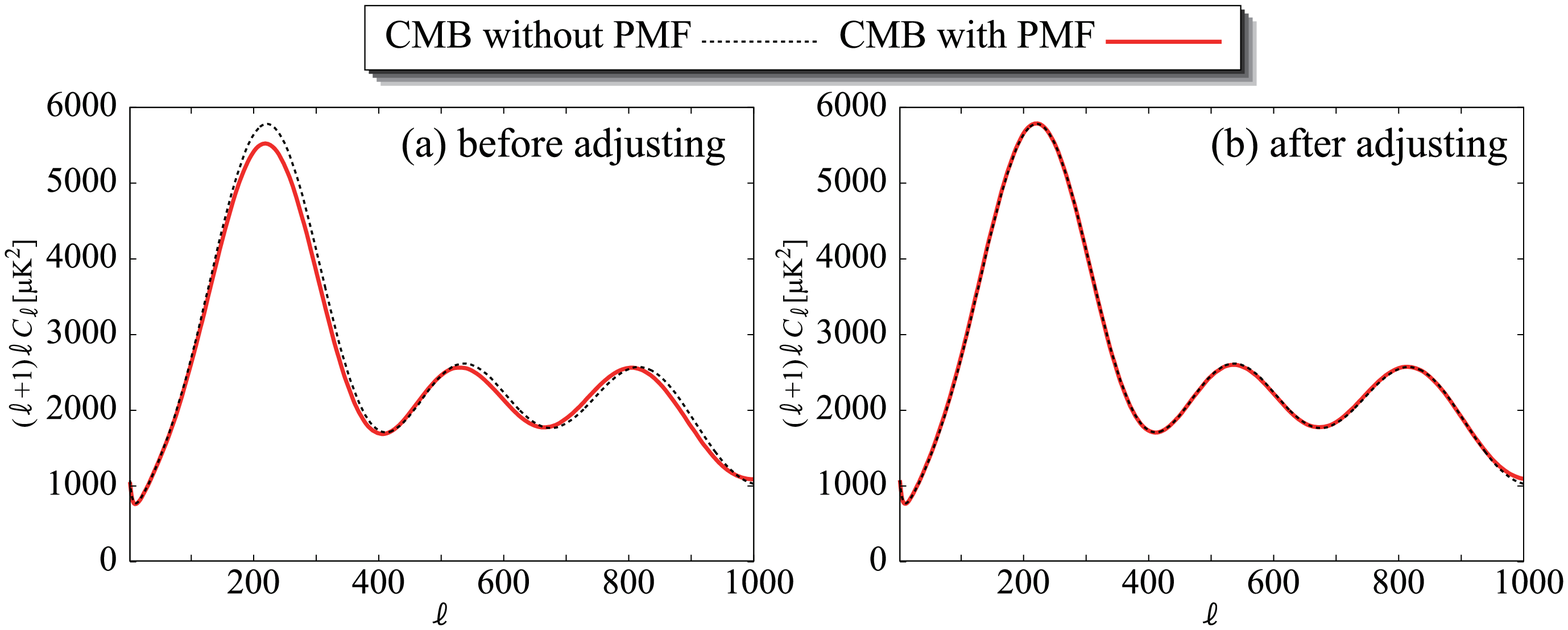}
\caption{\label{fig5}
The contribution of the PMF and the standard cosmological parameters to the CMB. The dotted curves are the theoretical results without PMF effects.
The bold curves are the theoretical results with PMF effects of 
$(n_\mathrm{B},~B_\lambda,~k_\mathrm{max},~\rho_\mathrm{MF}/\rho_\gamma)~=~(-2.0,~10~\mathrm{nG}, ~1500~\mathrm{Mpc}^{-1},~0.0160)$.
The standard cosmological parameters of all curves in this figure except the bold curves of panel (b) are the WMAP nine year best-fit parameters in $\Lambda$CDM and the tensor mode\cite{WMAP_9yr_Arxiv}.These standard cosmological parameters are $(\Omega_b, ~\Omega_\mathrm{CDM},~n_s,~ 10^9\Delta^2_R, ~H_0, ~\tau, r) = (0.0442, ~0.210,~0.992,~2.26,~72.6,~0.091,~0.38)$.
The different standard cosmological parameters of the bold curve in the right panel (b) are the baryon density and the CDM density. These parameter values are $(\Omega_b, ~\Omega_\mathrm{CDM}) = (0.0461, ~0.195)$.} 
\end{figure}

Finally, we discuss degeneracies between the background PMF, the baryon, and CDM.
Considering the fundamental understanding of the baryon and matter-density effects on the CMB \cite{Dodelson:2003booka}, 
we expect a positive correlation between $\Omega_\mathrm{b}$ and the background PMF, and a negative correlation between $\Omega_\mathrm{CDM}$ and the background PMF.
In fact, the affected CMB by the PMF can be adjusted by changing $\Omega_\mathrm{b}$ and $\Omega_\mathrm{CDM}$, as shown in Fig. \ref{fig5}.
In previous constraints on the PMF without the background effects, the degeneracy between the PMF and the standard cosmological parameters is negligibly small. If $n_\mathrm{B}$ and $k_\mathrm{max}$ are sufficiently small, the $\rho_\mathrm{MF}$ is too small to affect the CMB, and the previous result is no problem. 
However, a lot remains to be established about the PMF; it is too early to discuss the PMF effects and generation mechanisms in such a narrow parameter range.
To understand the PMF correctly, we should constrain the background PMF and the standard cosmological parameters simultaneously.
\section{Summary}
We consider the background PMF effects on the CMB for the first time.
The background PMF increases the sound speed of the tightly coupled photon-baryon fluid and causes a stronger decaying potential. The overall effect of the background PMF changes amplitudes of the CMB around the first peak more strongly than in smaller scales. Since the observational result of the CMB around the first peak is much better than at higher peaks, we expect that the PMF of bigger $n_\mathrm{B}$'s can be constrained more strongly.
We report the case in which $k_\mathrm{max}$ is assumed to be a free parameter. The energy density of the background PMF is dependent on $k_\mathrm{max}$, and $k_\mathrm{max}$ is dependent on a PMF generation model. Hence, if one determines the $k_\mathrm{max}$ by constraining the magnetic energy density from the CMB, we obtain information about the PMF generation mechanisms. We also discuss the constraint on the $k_\mathrm{max}$ as being an examination for the nonlinear evolution of the PMF.

Finally, we discuss the possibility of degeneracies between the background PMF, the baryon, and the CDM. 

If we promote the effects of the background PMF on the CMB, and constrain them by the latest and future observations, it will permit the development of better studies for the generation and evolution of the PMF and provide new insight into the early Universe with the PMF.

\begin{acknowledgments}
This work has been supported in part by Grants-in-Aid for Scientific Research 
(Grant No. 25871055) of the Ministry of Education, Culture, Sports,
Science and Technology of Japan.
We are grateful to Yolande McLean for improving the English in this paper.
\end{acknowledgments}
\appendix
\section{\label{sec:app1} A minimum scale of the PMF}
In this appendix, we mention briefly how to estimate a minimum scale of the PMF in the early Universe.
The electrical resistance in the early Universe is defined by
\begin{eqnarray}
\omega_e
\equiv
\frac{1}{\sigma}
     =
  \frac{m_e}{n_e e^2}
   c n_\gamma\sigma_\mathrm{T}
=
  \frac{c \sigma_\mathrm{T} m_e }{\eta e^2},
 \label{sigma}
\end{eqnarray}
where $\sigma$ is the electric conductivity, 
$m_e$ is the electron mass, 
$n_e$ is the electron number density, 
$e$ is the charge of an electron,
$c$ is the speed of light, 
$n_\gamma$ is the photon number density, 
$\sigma_\mathrm{T}$ is the Thomson scattering cross section, 
and $\eta$ is the baryon-to-photon ratio.
From Eq. (\ref{sigma}), the magnetic diffusivity is 
\begin{eqnarray}
\zeta \equiv \frac{c^2}{4\pi \sigma} = \frac{c^2\omega_e}{4\pi}. \label{zeta}
\end{eqnarray}
Since statistically average motions of fluids are assumed to be negligibly small in the early Universe, the induction equation from Eq. (\ref{zeta}), Ohm's law,  and Maxwell's equations is \cite{Dendy:1990booka}
\begin{eqnarray}
\frac{\partial \mbi{B}}{\partial t}=\zeta\nabla^2 \mbi{B}.\label{eq_MO} 
\end{eqnarray}
This equation indicates the magnetic field dissipating, and that the magnetic field dissipates rapidly with time and cannot survive on the scale length 
\begin{eqnarray}
L <
L_\mathrm{FI}(t)
\equiv 
\sqrt{\zeta t_\mathrm{age}}
=
7.5046\times
  10^{-2}
\frac{\mathrm{cm}}{\mathrm{sec}^{\frac{1}{2}}}
\left(
    \frac{t_\mathrm{age}}{\eta}
\right)^{\frac{1}{2}}
\label{eq:min}
\end{eqnarray}
\cite{Grasso:2000wj,Dendy:1990booka}, 
where $t_\mathrm{age}$ is the Universe's age.
The magnetic field on the scale length $L \ll L_\mathrm{FI}(T)$  is also difficult to produce.
On the other hand, the magnetic field on a scale length much bigger than $L_\mathrm{FI}$ is difficult to dissipate by time $t_\mathrm{age}$, and such a magnetic field is "frozen in" in the dominant fluids \cite{Dendy:1990booka}.
For example, from Eq.(\ref{eq:min}), the comoving minimum scale length of the magnetic field at last scattering is of the order of $10^{-11}$ Mpc.
\section{\label{sec:app2} A background energy density of PMF in the Universe}
In this appendix, we derive the background energy density of the power-law PMF.
A two-point correlation function of the PMF strength \cite{Yamazaki:2006mi,2010AdAst2010E..80Y,2012PhR...517..141Y} is defined by
\begin{eqnarray}
\left\langle 
	B^{i}(\mbi{k}) {B_{j}}^*(\mbi{k}')
\right\rangle 
	&=& 
	(2\pi)^3P_\mathrm{[PMF]}(k)P^{i}_j(k)\delta(\mbi{k}-\mbi{k}')~,
\label{eq:two_point_correlations_of_PMF}
\end{eqnarray}
where
\begin{eqnarray}
P^{i}_j(k)&=&
	\delta^{i}_j-\frac{k{}^{i}k{}_{j}}{k{}^2} \label{project_tensor}
\end{eqnarray}
and
\begin{eqnarray}
P_\mathrm{[PMF]}(k) = Ak^{n_\mathrm{B}}. \label{eq_power1}
\end{eqnarray}
We use the convention for the Fourier transform as
\begin{eqnarray}
f(\mbi{k}) = \int \mathrm{exp} (i\mbi{k} \cdot \mbi{x}) F(\mbi{x}) d^3 x .
\end{eqnarray}
Equation (\ref{eq:two_point_correlations_of_PMF}) gives 
\begin{eqnarray}
\left\langle 
	B^{i}(\mbi{k}) {B_{i}}^*(\mbi{k}')
\right\rangle 
&=& 2(2\pi)^3P_\mathrm{[PMF]}(k)\delta(\mbi{k}-\mbi{k}')
\nonumber \\
&=& 2(2\pi)^3 A k^{n_\mathrm{B}}\delta(\mbi{k}-\mbi{k}').
\label{eq:two_point_correlations_of_PMF2}
\end{eqnarray}

Next, we shall derive $A$.
We define
\begin{eqnarray}
\left.
	\left\langle
		B^{i}(\mbi{x}) {B_i}(\mbi{x})
	\right\rangle
\right|_\lambda
= B_\lambda^2.
\label{eq_real}
\end{eqnarray}
where $\lambda$ is a comoving scale for a Gaussian sphere on the present, and $B_\lambda$ is a comoving strength of PMF, and it is scaled to the present value on $\lambda$.
From Eqs.~(\ref{eq:two_point_correlations_of_PMF})~-~(\ref{eq_power1})~and~(\ref{eq_real}),
\begin{eqnarray}
\left.
	\left\langle
		B^{i}(\mbi{x}) {B_i}(\mbi{x})
	\right\rangle
\right|_\lambda
&=&
B_\lambda^2
\nonumber \\
&=&
	\frac{1}{(2\pi)^6}
	\int d^3 k 
	\int d^3 k'
		\exp(-i\mbi{x}\cdot\mbi{k}+i\mbi{x}\cdot\mbi{k}') 
\nonumber \\
&&\times 
		\left\langle
			B^{i}(\mbi{k}) {B_i^{\ast}}(\mbi{k}')
		\right\rangle
\times |W_{\lambda}^2(k)|,\label{eq_real2}
\end{eqnarray}
where $W_\lambda(k)$ is a Gauss window function as $W_\lambda(k) = \exp(-\lambda^2k^2/2)$. 
So we finally have 
\begin{eqnarray}
A &=& 
B^2_\lambda
\frac{(2\pi)^2}{4}
\left(
   \int dk k^{n_\mathrm{B}+2} \exp(-\lambda^2k^2)
\right)^{-1}
\nonumber \\
&=&
B^2_\lambda
\frac{(2\pi)^2}{2}
\frac{
\lambda^{n_\mathrm{B}+3}
}
{
   \Gamma
   \left(
      \frac{n_\mathrm{B}+3}{2}
   \right)
}
\nonumber \\
&=&
B^2_\lambda
\frac{(2\pi)^{n_\mathrm{B}+5}}{2}
\frac{1}
{
   k_\mathrm{[PMF]}^{n_\mathrm{B}+3}
   \Gamma
   \left(
      \frac{n_\mathrm{B}+3}{2}
   \right)
},
\label{eq_A}
\end{eqnarray}
where $\Gamma(x)$ is the gamma function and $\lambda=2\pi/k_\mathrm{[PMF]}$.
Substituting this  into Eq. (\ref{eq_power1}) leads to 
\begin{eqnarray}
P_\mathrm{[PMF]}(k)
= 
\frac{
(2\pi)^2 B^2_\lambda
\lambda^{n_\mathrm{B}+3}
}
{
   2
   \Gamma
   \left(
      \frac{n_\mathrm{B}+3}{2}
   \right)
}
k^{n_\mathrm{B}}.
\label{eq:two_point_correlations_of_PMF3}
\end{eqnarray}
From Eqs~(\ref{eq:two_point_correlations_of_PMF})-(\ref{eq:two_point_correlations_of_PMF2})~and~(\ref{eq:two_point_correlations_of_PMF3}),
the PMF energy density is derived by
\begin{eqnarray}
\rho_\mathrm{MF}
 &=& 
\frac{\langle B^2 \rangle}{8\pi}
 = 
\frac{2}{8\pi}
\int^{k_\mathrm{[max]}}_{k_\mathrm{[min]}}
\frac{dk}{k}
\frac{k^3}{2\pi^2}
P_\mathrm{[PMF]}(k)
\nonumber \\ 
 &=& 
\frac{2}{8\pi}
\int^{k_\mathrm{[max]}}_{k_\mathrm{[min]}}
\frac{dk}{k}
\frac{k^3}{2\pi^2}
\frac{
(2\pi)^2 B^2_\lambda
\lambda^{n_\mathrm{B}+3}
}
{  2
   \Gamma
   \left(
      \frac{n_\mathrm{B}+3}{2}
   \right)
}
k^{n_\mathrm{B}}
\nonumber \\ 
 &=& 
\frac{1}{8\pi}
\frac{
B^2_\lambda
}
{
   \Gamma
   \left(
      \frac{n_\mathrm{B}+5}{2}
   \right)
}
\left[
(\lambda k_\mathrm{[max]})^{n_\mathrm{B}+3}
-
(\lambda k_\mathrm{[min]})^{n_\mathrm{B}+3}
\right].
\nonumber \\
\label{eq:PL_PMF_energy_density}
\end{eqnarray}
Here $k_\mathrm{[max]}$ and $k_\mathrm{[min]}$ are the maximum and
minimum wave numbers, respectively. They are dependent on PMF generation models.
 The main goal of this study is to research the effects of the PMF energy density on the CMB and to discuss the degeneracy between the PL-PMF parameters and distribution models of the PMF. In order to effectively proceed with such research and discussions, from Eq.~(\ref{eq:PL_PMF_energy_density}), the scale-invariant (SI) strength of the PMF is defined by 
\begin{eqnarray}
\sqrt{\rho_\mathrm{MF}}  \propto B_\mathrm{SI}
\equiv
B_\lambda
\sqrt{
\frac{
\left[
(\lambda k_\mathrm{[max]})^{n_\mathrm{B}+3}
-
(\lambda k_\mathrm{[min]})^{n_\mathrm{B}+3}
\right]
}
{
   \Gamma
   \left(
      \frac{n_\mathrm{B}+5}{2}
   \right)
}
}
,
\end{eqnarray}
and 
\begin{eqnarray}
B_\lambda (n_\mathrm{B}, k_\mathrm{[max]}, k_\mathrm{[min]}) = 
B_\mathrm{SI}
   \sqrt{
   \frac{
   \Gamma \left(\frac{n_\mathrm{B}+5}{2}\right)
   }
   {
   \left(
      k_\mathrm{[max]} ^{n_\mathrm{B}+3}
     -
      k_\mathrm{[min]} ^{n_\mathrm{B}+3}
   \right)
   \lambda^{n_\mathrm{B}+3}
}
},
\nonumber \\
\label{eq:field_strength_for_elements}
\end{eqnarray}
where $B^2_\mathrm{SI}$ is directly proportional to the PMF energy,
and not dependent on other PMF parameters.
Therefore, these formulations are useful for directly understanding the PL-PMF energy density effects on the CMB, and also make it relatively easy to discuss the degeneracy of the PL-PMF parameters.
\bibliographystyle{apsrev}

\end{document}